\documentclass[twocolumn,aps]{revtex4}

\usepackage{graphicx}
\usepackage{dcolumn}
\usepackage{bm}

\begin{document}

\preprint{APS/123-QED}

\title{Exotic superconductivity in the coexistent phase of antiferromagnetism and superconductivity in CeCu$_2$(Si$_{0.98}$Ge$_{0.02}$)$_2$ : A Cu-NQR study under hydrostatic pressure}

\author{Y.~Kawasaki}
\altaffiliation[Present address : ]{Department of Physics, Faculty of Engineering, Tokushima University, Tokushima 770-8506, Japan}%
\author{K.~Ishida}%
\altaffiliation[Present address : ]{Department of Physics, Graduate School of Science, Kyoto University, Kyoto 606-8502, Japan}%
\author{S.~Kawasaki}%
\author{T.~Mito}%
\altaffiliation[Present address : ]{Department of Physics, Faculty of Science, Kobe University, Hyogo 657-8501, Japan}%

\author{G.-q.~Zheng}%
\author{Y.~Kitaoka}%
\affiliation{Department of Physical Science, Graduate School of Engineering Science, Osaka University, Toyonaka, Osaka 560-8531, Japan}
\author{C.~Geibel}%
\author{F.~Steglich}%
\affiliation{Max-Planck Institute for Chemical Physics of Solids, D-01187 Dresden, Germany}

\date{\today}

\begin{abstract}
We report a pressure ($P$) effect on CeCu$_2$(Si$_{0.98}$Ge$_{0.02}$)$_2$ where an antiferromagnetic (AFM) order at $T_N \sim$ 0.75 K coexists with superconductivity below $T_c \sim$ 0.4 K\@.
At pressures exceeding $P = 0.19$ GPa, the AFM order is suppressed, which demonstrates that the sudden emergence of AFM order due to the Ge doping is ascribed to the intrinsic lattice expansion.
The exotic superconductivity at $P = 0$ GPa is found to evolve into a typical heavy-fermion one with a line-node gap above $P = 0.91$ GPa\@.
We highlight that the anomalous enhancement in nuclear spin-lattice relaxation rate $1/T_1$ that follows a $T_1T$ = const.\ behavior well below $T_c$ at $P$ = 0 GPa is characterized by the persistence of low-lying magnetic excitations, which may be inherent to the coexistent state of antiferromagnetism and superconductivity.
\end{abstract}
\maketitle
%
\section{Introduction}

\sloppy
In the last decade, superconductivity has been discovered in several heavy-fermion (HF) antiferromagnets around a quantum critical point (QCP) where the N\'eel temperature $T_N$ is suppressed to zero under pressure ($P$) \cite{jaccard92,movshovich96,mathur98,hegger00}.
In these $P$-induced superconductors, the most promising interaction leading to the formation of Cooper pairs is magnetic in origin, presumably due to the closeness to an antiferromagnetic (AFM) phase.
However, the nature of superconductivity and magnetism is still unclear when the superconductivity appears very close to the counterpart AFM phase.
Therefore, in light of an exotic interplay between these phases, unconventional electronic and magnetic properties around QCP have attracted much attention and a lot of experimental and theoretical works are being extensively made.

In these experimental studies, QCP is realized by applying a critical pressure ($P_c$) to HF antiferromagnets such as CeCu$_2$Ge$_2$, CeRh$_2$Si$_2$, CeIn$_3$, CeRhIn$_5$, and so on.
Above $P_c$ where the AFM order is suppressed, unconventional HF superconducting (SC) nature is indicated from recent studies by means of nuclear-quadrupole-resonance (NQR) measurements on these compounds.
The nuclear spin-lattice relaxation rate $1/T_1$ in CeRhIn$_5$ is shown to follow a $T^3$ behavior below $T_c \sim$ 2.1 K at $P = 2.1$ GPa above $P_c \sim$ 1.6 GPa, which indicates the presence of line-node SC gap at the Fermi surface \cite{mito01,kohori00}.
As for CeIn$_3$, the absence of coherence peak in $1/T_1$ just below $T_c \sim 0.1$ K points to an unconventional type of superconductivity at $P = 2.65$ GPa above $P_c \sim$ 2.55 GPa\@.\cite{shinji02}
Thus, these observations of unconventional superconductivity above $P_c$ are consistent with the model that magnetic interactions play an important role.

On the other hand, the magnetic and SC properties {\it just below} $P_c$ may be much more exotic than those in the previous examples, since superconductivity could appear in the AFM phase below $T_c$\@.
The situation reminds us of the firstly-discovered HF superconductor CeCu$_2$Si$_2$ located just at the border of the AFM phase at ambient pressure \cite{ishida99}.
This was evidenced by various magnetic anomalies observed above $T_c$ \cite{nakamura,steglich79} and by the fact that the magnetic phase $A$ appears when the superconductivity is suppressed by a magnetic field \cite{bruls94}.
Furthermore, the transport, thermodynamic, and NQR measurements have consistently shown that nominally off-tuned Ce$_{0.99}$Cu$_{2.02}$Si$_2$ is located near $P_c$ and crosses QCP by applying a minute pressure of 0.2 GPa\cite{gegenwart98,kawasaki01}.


In our previous papers,\cite{kawasaki01,kawasaki02} the magnetic and SC properties in CeCu$_2$Si$_2$ were investigated around the QCP as functions of pressure for Ce$_{0.99}$Cu$_{2.02}$Si$_2$ and of Ge content $x$ for CeCu$_2$(Si$_{1-x}$Ge$_x$)$_2$ by the Cu-NQR measurements.
Figure 1 shows the phase diagram referred from the literature.\cite{kawasaki02}
Here, $T_F^*$ is an effective Fermi temperature below which $1/T_1T$ stays constant and $T_m$ is a temperature below which the slowly fluctuating AFM waves start to develop.
Note that a primary effect of the Ge doping is assumed to expand the lattice\cite{trovarelli97} and that its chemical pressure is $-0.076$ GPa per 1\% Ge doping as suggested from the $P$ variation of Cu-NQR frequency $\nu_Q$ in CeCu$_2$Ge$_2$ and CeCu$_2$Si$_2$ \cite{kitaoka95}.

\begin{figure}
\begin{center}
\includegraphics[scale=0.45]{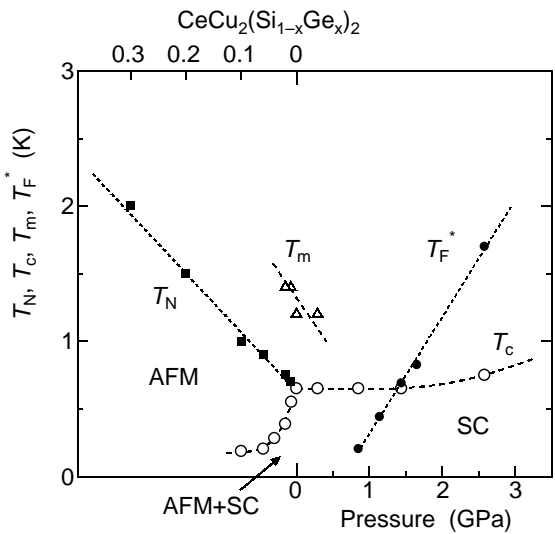}
\caption{The combined phase diagram of antiferromagnetism (AFM) and superconductivity (SC) for CeCu$_2$(Si$_{1-x}$Ge$_{x}$)$_2$ and for Ce$_{0.99}$Cu$_{2.02}$Si$_2$ under pressure.
$T_N$ and $T_c$ are the respective transition temperature of AFM and SC\@.
Also shown are $T_m$ below which the slowly fluctuating AFM waves develop and $T_F^*$ below which $1/T_1T$ becomes const.
}
\label{fig:2}
\end{center}
\end{figure}
Ce$_{0.99}$Cu$_{2.02}$Si$_2$ is just at the border of the AFM phase at ambient pressure.
In the normal state, the slowly fluctuating AFM waves propagate over a long range without any trace of AFM order below $T_m \sim 1.2$ K\@.
The most striking feature may be in the exotic SC state below $T_c \sim 0.65$ K, where low-lying magnetic excitations seem to remain active, since $1/T_1$ shows a rapid decrease below $T_c$, followed by the large enhancement in $1/T_1T$ well below $T_c$\@.
With increasing $P$, the slowly fluctuating AFM waves are markedly suppressed and, concomitantly, the exotic superconductivity evolves into a typical HF-SC one characterized by the relation of $1/T_1\propto T^3$ above 0.85 GPa\@.


By substituting only 1\% Ge, the AFM order emerges below $T_N \sim$ 0.7 K, followed by the SC transition at $T_c \sim$ 0.5 K \@.
Nevertheless, $1/T_1$ does not show distinct reduction at $T_c$, but follows a $1/T_1T$ = const.\ behavior well below $T_c$ as observed in Ce$_{0.99}$Cu$_{2.02}$Si$_2$\@.
It seems that the exotic superconductivity accompanied by the slowly fluctuating AFM waves coexists with the AFM order below $T_c$\@.
As the Ge content increases, $T_N$ is progressively increased, while $T_c$ is steeply decreased.
As a result of the suppression of the slowly fluctuating AFM waves in the samples more than $x = 0.06$, the magnetic properties above $T_N$ progressively change to those in a localized regime as observed in CeCu$_2$Ge$_2$\@.\cite{kitaoka95}
It was also proposed that these exotic SC natures around the border of the AFM phase could be accounted for on the basis of SO(5) theory that unifies superconductivity and antiferromagnetism \cite{kitaoka01,zhang97}.

In order to get further insight into the exotic SC state near the AFM phase and to clarify the role of Ge doping, we have carried out Cu-NQR study on CeCu$_2$(Si$_{0.98}$Ge$_{0.02}$)$_2$ ($T_N \sim$ 0.75 K, $T_c \sim$ 0.4 K) under pressure.
The Cu-NQR measurements have shown that the AFM order in CeCu$_2$(Si$_{0.98}$Ge$_{0.02}$)$_2$ is suppressed at pressures exceeding 0.19 GPa and that a typical HF-SC behavior recovers at $P = $ 0.91 GPa\@.
These results demonstrate that the primary effect of the Ge doping is to expand the lattice, leading to the sudden emergence of the AFM order.
We stress that the anomalous enhancement of $1/T_1$ below $T_c$ at $P = 0$ GPa is ascribed not to impurity effect but to the persistence of low-lying magnetic excitations.

\section{Experimental}

The same polycrystalline sample of CeCu$_2$(Si$_{0.98}$Ge$_{0.02}$)$_2$ as in the previous work\cite{kawasaki02} was used.
The detailed preparation method of the sample has been already reported in the literature \cite{trovarelli97}.
Hydrostatic pressure was applied by utilizing a CuBe piston-cylinder cell with the Daphne oil 7373 as a pressure-transmitting medium.
To calibrate a value of $P$ at a sample position at low temperatures, the $P$-induced variation of $T_c$ of Sn was measured by a high-frequency ac-$\chi$ measurement using in-situ NQR coil.
As $P$ increases, $\nu_Q$ increases linearly.
This result suggests that an increase in the lattice density gives rise to a linear increase of the electric field gradient at the Cu site.
The rate of the increase in $\nu_Q$, $d\nu_Q/dp = 9.52$ Hz/bar is in good agreement with those for Ce$_{0.99}$Cu$_{2.02}$Si$_2$ and CeCu$_2$Ge$_2$ \cite{kitaoka95,kawasaki01}.
The gradual increase of the NQR linewidth with increasing $P$ assures that a possible distribution of $P$ is as small as $\pm$ 0.08 GPa at $P = 0.91$ GPa\@.

\section{Results and Discussion}

The AFM order in CeCu$_2$(Si$_{0.98}$Ge$_{0.02}$)$_2$ at $P$ = 0 GPa is evidenced by the rapid increase of the NQR linewidth and the peak in $1/T_1$ at $T_N \sim$ 0.75 K, as reported previously \cite{kawasaki02,kitaoka01}.
Figure 2(a) and the main panel of Fig.~3 show the $^{63,65}$Cu-NQR spectra at several temperatures and the $T$ dependence of $1/T_1$ at $P$ = 0 GPa (closed circles), respectively.
Here, $1/T_1$ is uniquely determined above $T_N$, but not below $T_N$\@.
This distribution in $1/T_1$ is presumably associated with the distribution in the internal field at the Cu site due to the incommensurate magnetic structure in the ordered state.\cite{kawasaki02}
Only main components in $1/T_1$ are shown below $T_N$ for simplicity.
As seen in Fig.~2(a), the increase in the NQR linewidth indicates an appearance of the internal field at the Cu site below $T_N$\@. Correspondingly, $1/T_1$ shows a clear peak at the same temperature, reflecting a critical slowing down as approaching the AFM transition.

\begin{figure}
\begin{center}
\includegraphics[scale=0.45]{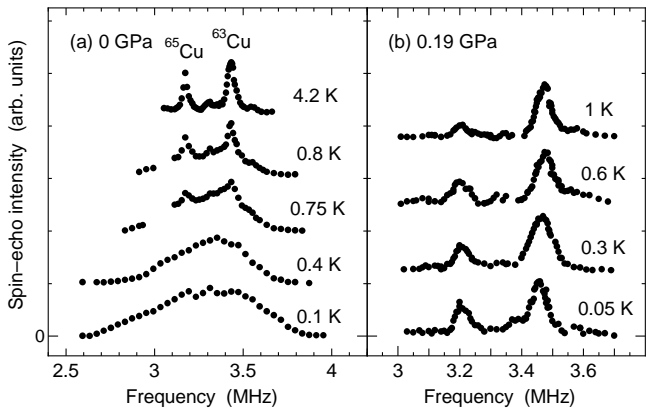}
\caption{$^{63,65}$Cu-NQR spectra at several temperatures at (a) $P$ = 0 GPa and at (b) 0.19 GPa.}
\label{fig:2}
\end{center}
\end{figure}
Now, we turn to the present results under $P$\@.
The $T$ dependence of NQR spectrum at 0.19 GPa is shown in Fig.~2(b)\@.
Apparently, any broadening of the NQR spectrum is not observed down to 0.05 K at $P$ = 0.19 GPa, in contrast to the significant broadening of NQR linewidth below $T_N$ at $P$ = 0 GPa\@.
This result indicates that the AFM order is suppressed by applying the minute pressure of 0.19 GPa, which is also corroborated by the $T$ dependence of $1/T_1$ as shown later.
Note that $\nu_Q$ at 0.19 GPa decreases slightly by 0.02 MHz as $T$ decreases below $T_c\sim 0.45$ K\@.
By contrast, at 0.91 GPa where the superconductivity shows the typical HF-SC behavior as shown later, such a shift in $\nu_Q$ is not observed.
Therefore, the slight decrease in $\nu_Q$ below $T_c$ at 0.19 GPa may be related to the exotic SC state due to the closeness to the AFM phase.

In Fig.~3 are shown the $T$ dependencies of $1/T_1$ at $P$ = 0 GPa (closed circles), 0.56 GPa (open circles) and 0.91 GPa (closed squares).
Here, the data at $P$ = 0.19 GPa are not shown, since they are nearly equivalent to those at $P$ = 0.56 GPa\@.
The distribution in $1/T_1$ observed below $T_N$ at $P = 0$ GPa is no longer appreciable at these elevated pressures.
In the entire $T$ range, $1/T_1$ is suppressed with increasing $P$, evidencing that the low-energy component of spin fluctuations (SF's) is forced to shift to a high-energy range.
As expected from the fact that the AFM order is already suppressed at pressures exceeding 0.19 GPa, any trace of anomaly associated with it is not observed at all down to $T_c \sim$ 0.45 K at $P$ = 0.56 GPa and 0.91 GPa\@.
It is, therefore, considered that the AFM order in CeCu$_2$(Si$_{0.98}$Ge$_{0.02}$)$_2$ is not triggered by some disorder effect but by the intrinsic lattice expansion due to the Ge doping.

\begin{figure}
\begin{center}
\includegraphics[scale=0.43]{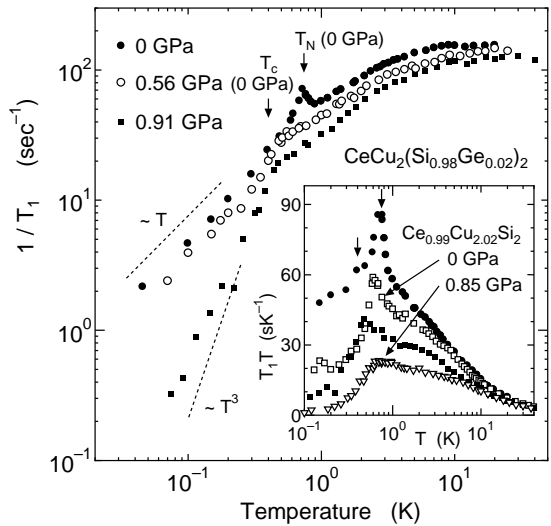}
\caption{$T$ dependence of $1/T_1$ of CeCu$_2$(Si$_{0.98}$Ge$_{0.02}$)$_2$ at several pressures.
Inset shows the $T$ dependence of $1/T_1T$ of CeCu$_2$(Si$_{0.98}$Ge$_{0.02}$)$_2$ at $P$ = 0 GPa (closed circles) and 0.91 GPa (closed squares) and those of Ce$_{0.99}$Cu$_{2.02}$Si$_2$ at $P$ = 0 (open squares) and 0.85 GPa (open triangles).
Arrows indicate $T_N \sim 0.75$ K and $T_c \sim$ 0.4 K at $P = 0$ GPa for CeCu$_2$(Si$_{0.98}$Ge$_{0.02}$)$_2$\@.}
\label{fig:2}
\end{center}
\end{figure}
In order to demonstrate a systematic evolution of SF's in the paramagnetic state, in the inset of Fig.~3 are shown the $T$ dependencies of $1/T_1T$ in CeCu$_2$(Si$_{0.98}$Ge$_{0.02}$)$_2$ at $P$ = 0 GPa (closed circles) and 0.91 GPa (closed squares) along with the results in Ce$_{0.99}$Cu$_{2.02}$Si$_2$ at $P$ = 0 GPa (open squares) and 0.85 GPa (open triangles).\cite{kawasaki01}
The $1/T_1T$ result in CeCu$_2$(Si$_{0.98}$Ge$_{0.02}$)$_2$ at $P$ = 0 GPa is well explained by the SF's theory for weakly AFM itinerant magnets in $T_c < T < 1.5$ K around $T_N \sim$ 0.75 K as reported previously \cite{kitaoka01,kawasaki02}.
The good agreement between the experiment and the calculation indicates a long-range nature of the AFM ordering in the itinerant regime.
At $P =$ 0.91 GPa, $1/T_1T$, which probes the development of magnetic excitations, is suppressed and resembles a behavior that would be expected at an intermediate pressure between $P$ = 0 GPa and 0.85 GPa for Ce$_{0.99}$Cu$_{2.02}$Si$_2$\@.
This reveals that the Ge doping works as applying a negative chemical pressure to expand the lattice of Ce$_{0.99}$Cu$_{2.02}$Si$_2$\@.

Figure 4 indicates the $T$ dependence of NQR intensity multiplied by temperature $I(T)\times T$ in CeCu$_2$(Si$_{0.98}$Ge$_{0.02}$)$_2$ at $P =$ 0, 0.19, 0.56, and 0.91 GPa\@.
Here, $I(T)$ normalized by the value at 4.2 K is an integrated intensity over frequencies where the NQR spectrum was observed.
Since the NQR intensity depends on a pulse interval $\tau$ between two pulses for the spin-echo measurement, $I(\tau=0)$ is evaluated through the relation of $I(\tau) = I(\tau = 0)\exp(-2\tau/T_2)$, where $1/T_2$ is the spin-echo decay rate.
Note that $I(T)\times T$ stays constant generally, if $T_1$ and/or $T_2$ range in the observable time window that is typically more than several microseconds.
Therefore, the distinct reduction in $I(T)\times T$ upon cooling is ascribed to the development of the slowly fluctuating AFM waves, since it leads to an extraordinary short relaxation time of $\sim$ 0.14 $\mu$sec \cite{ishida99}.

The $I(T)\times T$ at $P = 0$ GPa decreases down to about 0.55 at $T_N \sim 0.75$ K upon cooling below $T_m \sim$ 1.2 K\@.
The reduction in $I(T)\times T$ stops around $T_N$, but does no longer recover with further decreasing $T$\@.
Note that the reduction in $I(T)\times T$ below $T_c \sim 0.4$ K is due to the SC diamagnetic shielding of rf field for the NQR measurement.
As $P$ increases, $T_m$ becomes smaller, in agreement with the result presented in the phase diagram of Fig.~1, and the reduction in $I(T)\times T$ in the normal state becomes moderate.
With further increasing $P$ up to 0.91 GPa, eventually, $I(T)\times T$ remains nearly constant down to $T_c \sim 0.45$ K, indicative of no anomaly related to the slowly fluctuating AFM waves.
This behavior resembles the result observed at pressures exceeding 0.85 GPa in Ce$_{0.99}$Cu$_{2.02}$Si$_2$\@.
These results also assure that the Ge substitution expands the lattice of Ce$_{0.99}$Cu$_{2.02}$Si$_2$\@.

\begin{figure}
\begin{center}
\includegraphics[scale=0.43]{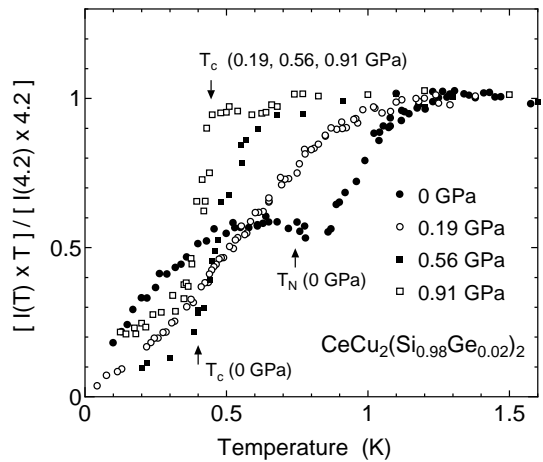}
\caption{$T$ dependence of $I(T)\times T$ at several pressures, where $I(T)$ is an NQR intensity normalized by the value at 4.2 K\@.}
\label{fig:2}
\end{center}
\end{figure}

In the previous paper,\cite{kawasaki02} we highlighted that the exotic superconductivity in Ce$_{0.99}$Cu$_{2.02}$Si$_2$ and CeCu$_2$(Si$_{0.98}$Ge$_{0.02}$)$_2$ at $P = 0$ GPa is characterized by the persistence of low-lying magnetic excitations.
It was argued that these excitations may be related to a collective mode in the coexistent phase of antiferromagnetism and superconductivity.
However, there remains a possibility that the potential scattering due to non-magnetic Ge impurities brings about a finite density of states at the Fermi surface, leading to a $T_1T =$ const.\ behavior below $T_c$ \cite{schmitt-rink86,hotta93}.
Here, we discuss an intimate $P$-induced evolution of low lying magnetic excitations in the SC state.
As seen in Fig.~3 and its inset, $1/T_1$ and $1/T_1T$ at $P =$ 0 GPa does not show a distinct reduction below $T_c$, but instead a $T_1T=$ const.\ behavior emerges well below $T_c$\@.
At $P$ = 0.56 GPa, the AFM order is absent, but the slowly fluctuating AFM waves develop in the normal state. It is noteworthy that the relation of $1/T_1\propto T$ is still valid below $T_c$, resembling the behavior for Ce$_{0.99}$Cu$_{2.02}$Si$_2$ at $P = 0$ GPa.\cite{kawasaki01,ishida99}
Remarkably at $P =$ 0.91 GPa, $1/T_1$ follows a relation of $1/T_1 \propto T^3$ below $T_c \sim$ 0.45 K, consistent with the line-node SC gap at the Fermi surface.
This typical HF-SC behavior in $1/T_1$ was observed in Ce$_{0.99}$Cu$_{2.02}$Si$_2$ at pressures exceeding 0.85 GPa as well \cite{kawasaki01}.
A small deviation from $1/T_1\propto T^3$ behavior at $P = 0.91$ GPa far below $T_c$ may be associated with an inevitable Ge-impurity effect for $d$-wave superconductors in general \cite{schmitt-rink86,hotta93}.
Therefore, it is considered that the unconventional SC characteristics at $P$ = 0 GPa and 0.56 GPa evolve into the typical HF-SC ones with the line-node gap at pressures exceeding 0.91 GPa\@.
Apparently, these results exclude a possible impurity effect as a primary cause for the $T_1T$ = const.\ behavior below $T_c$ at $P =$ 0 GPa\@.

While both the magnetic and SC properties undergo the marked change with increasing $P$, the value of $T_c \sim$ 0.4 K slightly increases up to 0.45 K at 0.19 GPa and saturates for the further increasing $P$\@.
Here, $T_c$'s are consistently determined by both the kink in $1/T_1$ and the onset of SC diamagnetism. The increment in $T_c$ is, however, smaller than the value that would be expected from the phase diagram in Fig.~1, where $T_c$ goes up to 0.65 K at the SC phase at higher pressures. This means that some impurity effect may be responsible for the reduction in $T_c$ due to the Ge doping.
In this sense, if the lattice were homogeneously expanded as for CeCu$_2$Si$_2$ without substituting Ge for Si, $T_c$ would be not so reduced and remain comparable to $\sim 0.65$ K\@.
It seems, therefore, that this exotic superconductivity could be rather robust against the appearance of AFM order, although the SC state evolves into the exotic SC state accompanied by low-energy magnetic excitations.
Note that $T_c$ decreases rapidly with the slight Ge substitution but remains finite with further increasing $x$ up to $x\sim 0.15$, as indicated in Fig.~1 and in the literature \cite{trovarelli97}.
This $x$ dependence of $T_c$ is apparently different from the behavior observed in usual $d$-wave superconductors with non-magnetic impurities.
It may be accounted for by the combined contributions of the impurity effect and of the unknown SC characteristics in the coexistent state of antiferromagnetism and superconductivity.

\section{Conclusion}

In conclusion, we have performed the Cu-NQR experiments under $P$ on CeCu$_2$(Si$_{0.98}$Ge$_{0.02}$)$_2$ where the AFM order below $T_N \sim 0.75$ K coexists with the superconductivity below $T_c \sim 0.4$ K\@.
The AFM order is suppressed by applying the minute pressure of 0.19 GPa\@.
As a result, the typical HF-SC behavior with the line-node gap recovers at $P$ = 0.91 GPa, demonstrating that the AFM order in CeCu$_2$(Si$_{0.98}$Ge$_{0.02}$)$_2$ is triggered by the lattice expansion caused by the Ge doping.
We highlight that the $1/T_1T$ = const.\ relation well below $T_c$ at $P = 0$ GPa is ascribed not to the impurity effect but to the persistence of low-lying magnetic excitations.
This feature may be inherent to the coexistent state of antiferromagnetism and superconductivity on a microscopic scale.

\section*{Acknowledgements}

This work was supported by the COE Research grant from MEXT of Japan (Grant No.\ 10CE2004).
One of the authors (Y.\ Kawasaki) was supported by the JSPS Research Fellowships for Young Scientists\@.

\end{document}